\documentstyle[12pt,epsf]{article}
\newcommand{\be}{\begin{equation}}
\newcommand{\bea}{\begin{eqnarray}}
\newcommand{\ee}{\end{equation}}
\newcommand{\eea}{\end{eqnarray}}

\begin{document}
\begin{titlepage}
%\thispagestyle{empty}
%\begin{flushright}
%UG-DFM-1/98 \\
%nucl-th/0008034
%\end{flushright}  
  \vspace*{5mm}

\vspace*{2cm}

\begin{center}
{\Large \bf Bethe-Salpeter Approach for
the $P_{33}$ Elastic Pion-Nucleon Scattering in Heavy Baryon Chiral
Perturbation Theory.}

\vspace{1.5cm}
{\large{\bf J. Nieves\footnote{email:jmnieves@ugr.es}
  and E. Ruiz Arriola\footnote{email:earriola@ugr.es}}}\\[2em]
Departamento de F\'{\i}sica Moderna, Universidad de Granada, 
E-18071 Granada, Spain.

\end{center}

\vspace{1cm}
\begin{abstract}
\vspace{1cm} Heavy Baryon Chiral Perturbation Theory (HBChPT) to
leading order provides a kernel to solve the Bethe-Salpeter equation
for the $P_{33}$ ($\Delta(1232)$-channel) $\pi-N$ system, in the
infinite nucleon mass limit. Crossed Born terms include, when iterated
within the Bethe-Salpeter equation, both {\it all} one- and {\it some}
two-pion intermediate states, hence preserving elastic unitarity below
the two-pion production threshold. This suggests searching for a
solution with the help of dispersion relations and suitable
subtraction constants, when all in-elasticities are explicitly
neglected. The solution allows for a successful description of the
experimental phase shift from threshold up to $\sqrt{s}=1500$ MeV in
terms of four subtraction constants. Next-to-leading order HBChPT
calculations are also used to estimate the unknown subtraction
constants which appear in the solution. Large discrepancies are
encountered which can be traced to the slow convergence rate of
HBChPT. 

\end{abstract}

\vspace*{1cm}
\centerline{\it PACS: 11.10.St;11.30.Rd; 11.80.Et; 13.75.Lb;
14.40.Cs; 14.40.Aq\\} 
\vspace*{1cm}
\centerline{\it Keywords: Bethe-Salpeter Equation, Heavy Baryon
Chiral Perturbation Theory,}
\centerline{\it Unitarity, $\pi N$-Scattering,
$\Delta$ Resonance, Dispersion Relations. }

\end{titlepage}

\newpage

\setcounter{page}{1}

\section{Introduction}

The study of meson-baryon scattering is one of the most challenging
topics of particle physics at low and intermediate energies. Firstly,
because there exist accurate experimental data and secondly because it
provides an excellent scenario where Chiral Symmetry Breaking (ChSB)
can be tested. Unlike the light pseudo-scalar meson--meson scattering,
the role played by ChSB in the pseudo-scalar meson--baryon scattering
case turns out to be more subtle and the construction of an Effective
Field Theory (EFT) to describe the latter processes becomes more
cumbersome. In the light meson--meson sector the expansion parameter
is $p^2/(4\pi f)^2$, being $p^\mu$ the four-momentum of the Goldstone
bosons~\cite{gl84} and $f$ the corresponding meson weak decay
constant. Since the baryon mass is already of the order of $4\pi f$,
such an expansion does not work in practice for the meson--baryon
sector, though some efforts were undertaken~\cite{gss88}. This
drawback was overcome by formulating a new consistent derivative
expansion for baryons in a chiral effective field
theory~\cite{jm91,BK92,BKM97}, called Heavy Baryon Chiral Perturbation
Theory (HBChPT). HBChPT is a reformulation of Chiral Perturbation
Theory (ChPT) with baryons where the baryon mass is shifted from the
propagator to the interaction couplings of the EFT~\footnote{This
program was firstly suggested, consistently formulated and
experimentally tested in the case of heavy quark ($Q=$ $b$, $c$)
physics~\protect\cite{iw89}.}. In this way higher orders are
suppressed either by powers of $p^2/(4\pi f)^2$ or by powers of $p/M$,
where $M$ is the baryon mass and $p$ is either the light pseudo-scalar
meson mass or the off-shellness of the baryon, respectively. Recently,
a fully relativistic formulation consistent with the power counting
proposed in HBChPT has been proposed~\cite{BL99}, but there is so far a
lack of practical applications.

Standard HBChPT has been applied to the $\pi N$ system several times
already in subsequently higher orders~\cite{BKM97,Mo98,FMS98,FM00} and
successful fits in the close to threshold region have been found. 
On the other hand HBChPT, as it is also the case for ChPT in the meson
sector, is unable to describe resonances, which clearly indicate the
appearance of non-perturbative chiral dynamics. Resonances, play a
crucial role in meson-baryon systems. For instance, the low and
intermediate energy $\pi N$ scattering is mostly dominated by the
$\Delta (1232)$ resonance, since it is close to threshold and its
coupling to pions and nucleons turns out to be quite
strong~\cite{ew88}. Hence, the $\Delta$ degrees of freedom are often
explicitly incorporated in any effective approach aiming at describing 
low and intermediate energy pion-nucleon dynamics to some degree of
accuracy~\cite{BKM95}-\cite{MO2000}.

Exact Unitarity plays a crucial role in the description of resonances,
since at the corresponding energy the scattering amplitude takes its
maximum possible value. HBChPT only restores unitarity perturbatively
and hence even the computation of phase-shifts becomes ambiguous
beyond perturbation theory. Actually, in previous works the
unitarization program of HBChPT for the lowest $S$ and $P$ partial
wave amplitudes obtained in has already been studied up to third order
in the chiral expansion, by using the Inverse Amplitude Method (IAM)
either in its most straightforward version~\cite{NP2000} or within an
improved IAM~\cite{NNPR2000}\footnote{The literature on unitarization
methods is vast. References in the present context can be
traced from Ref.~\protect{\cite{NNPR2000}}.}. We note here that since
the unitarity correction takes first place at NNLO in HBChPT one needs
to go to that order in order to unitarize \`a la IAM.

The aim of the present work is to show how a unitarization framework
based on the Bethe-Salpeter Equation (BSE)~\cite{BS51} and HBChPT to
lowest order yields to a very simple expression for the scattering
amplitude which preserves unitarity and provides a satisfactory
description of $\pi N$ elastic scattering data. Our work can be
regarded as a natural extension of the approach previously proposed in
Refs.~\cite{ej99} and~\cite{ej2000} for $\pi \pi $ scattering to the
physics of the $\pi N$ system.

We have concentrated our attention into the $P_{33}$ ($L_{2I2J}$)
channel, and show that the existence of the $\Delta$ resonance can be
understood in terms of the $\pi N$ interactions constrained to HBChPT.
Hence there is no particular need to include the $\Delta$-degrees of
freedom explicitly in order to provide a successful description of the
experimentally observed phase-shifts. The extension to other partial
waves as well as higher order contributions in HBChPT will be analyzed
and presented elsewhere~\cite{ej2001}.

\section{The Bethe-Salpeter Equation and the 
Elastic $P_{33}$ $ \pi N$ Amplitude in HBChPT.}

The BSE for the elastic $\pi N$ scattering amplitude ($T^I$) in a
given isospin channel ,$I$, reads
\begin{figure}
\vspace{-9cm}
\hbox to\hsize{\hfill\epsfxsize=0.75\hsize
\epsffile[52 35 513 507]{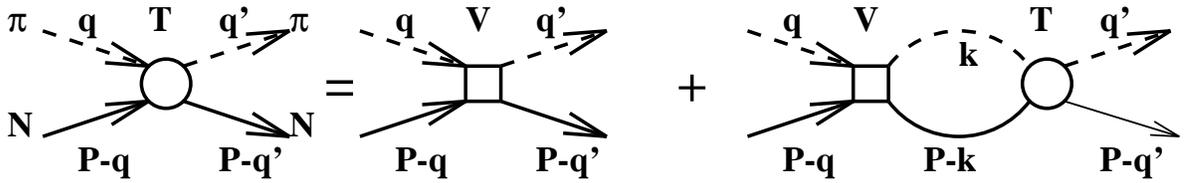}\hfill}
\vspace{-1.5cm}
\caption{\footnotesize Diagrammatic representation of the BSE
equation. It is also 
sketched the used kinematics.}
\label{fig:kin}
\end{figure}
\begin{eqnarray}
t_P^I(q,q^\prime) &=& v_P^I(q,q^\prime) + {\rm
i}\int\frac{d^4 k}{(2\pi)^4} t_P^I(k,q^\prime) D(k) S(P-k)
v_P^I(q,k)\label{eq:bs} 
\end{eqnarray}
with
\begin{eqnarray}
T_P^I(q,q^\prime)_{\sigma,\sigma^\prime} 
&=& {\bar u}_{\sigma^\prime} (P-q^\prime) 
\,t_P^I(q,q^\prime)\, u_\sigma(P-q) \nonumber\\
V_P^I(q,q^\prime)_{\sigma,\sigma^\prime} 
&=& {\bar u}_{\sigma^\prime}(P-q^\prime)\, 
v_P^I(q,q^\prime)\, u_\sigma(P-q) 
\end{eqnarray}
where the kinematical variables are given in Fig.~\ref{fig:kin} and
 $V^I$, $D$ and $S$ are the two particle irreducible amputated Green
 function, pion and nucleon propagators respectively. $t^I$, $v^I$ and
 $u$ are matrices and a spinor~\footnote{We use the normalization
 ${\bar u }u = 2M$.}  in the nucleon Dirac space and finally,
 $\sigma,\sigma^\prime$ are nucleon spin indices (helicity, covariant
 spin, etc...).  The normalization of the amplitude $T^I$ is
 determined by its relation with the center or mass (CM) differential
 cross section, in the isospin channel $I$, and it is given by
\begin{eqnarray}
\frac{d\sigma}{d\Omega} \left (\vec{q^\prime}, \sigma^\prime \leftarrow 
\vec{q}, \sigma \right ) &=& \frac{1}{64\pi^2 s}
\left |\, T_P^I(q,q^\prime)_{\sigma \sigma^\prime} \, \right  |^2
\end{eqnarray}
where $s=P^2$. Rotational, parity and time reversal invariances ensures
for the on shell particles
\begin{eqnarray}
T_P^I(q,q^\prime )_{\sigma,\sigma^\prime } = -8\pi\sqrt{s} \left \{ {\cal
A}^I(s,\theta )\delta_{\sigma\sigma^\prime } + {\rm i}\, {\cal
B}^I(s,\theta ) \left ( \hat{n} \cdot \vec {\sigma} \right
)_{\sigma^\prime \sigma } \right \}
\end{eqnarray}
with $\theta$  the CM angle between the initial and final pion three
momentum and $\hat{n}$  a unit three-vector orthogonal to
$\vec{q}$ and $\vec{q^\prime}$. Partial waves, $f_{IL}^J(s)$, 
are related  to ${\cal A}, {\cal B}$ by 
\begin{eqnarray}
{\cal A}^I(s,\theta ) &=& \sum_L \left [ (L+1)\,\, f_{IL}^{L+\frac12}(s) + 
L\,\, f_{IL}^{L-\frac12}(s) \right ] P_L(\cos\theta) \nonumber\\
{\cal B}^I(s,\theta ) &=& -\sum_L \left [  f_{IL}^{L+\frac12}(s) -  
\,\,f_{IL}^{L-\frac12}(s) \right ] \frac{d P_L(\cos\theta)}{d\theta}
\end{eqnarray}
The phase of the  amplitude $T^I$ is such that the relation between the
partial wave amplitudes and phase-shifts is the usual one,
$f_{IL}^J(s) =  
e^{{\rm i}\,\, \delta_{IL}^J(s)}\sin
\delta_{IL}^J(s)/|\,\vec{q}\,|_{CM}$. Hence, on-shell unitarity
implies for $s \ge (m+M)^2$
\begin{equation}
{\rm Im}[f_{IL}^J(s)]^{-1} = -  |\,\vec{q}\,|_{CM} = -
\frac{\lambda^\frac12(s,M^2,m^2)}{2\sqrt s} \label{eq:uni_f}
\end{equation} 
with $M= 938.27$ MeV, $m = 139.57$ MeV, the nucleon and pion masses
respectively, and $\lambda(x,y,z) = x^2+y^2+z^2 -2xy-2xz-2yz$. 

Note that, to solve Eq.~(\ref{eq:bs}) both the off-shell {\it potential} and
amplitude are required.  Clearly, for the exact {\it potential} $V^I$
and propagators $D$ and $S$, the BSE provides an exact solution of the
scattering amplitude $T^I$~\cite{BS51}. Obviously an exact solution
for $T^I$ is not accessible, since $V^I$, $D$ and $S$ are not exactly
known. We propose  an expansion along  the lines of HBChPT both for
the exact {\it potential} ($V^I$) and the exact propagators.

The lowest order of the HBChPT expansion in the $P_{33}$ channel leads
to~\cite{Mo98,FMS98}
\begin{eqnarray}
V_P^{I=\frac32}(q,q^\prime)_{\sigma,\sigma^\prime} &\approx&
V_P^{I=\frac32}(q,q^\prime)_{\sigma,\sigma^\prime} |_{\rm crossed} + 
V_P^{I=\frac32}(q,q^\prime)_{\sigma,\sigma^\prime} |_{\rm contact}
\nonumber \\
&&\nonumber\\
V_P^{I=\frac32}(q,q^\prime)_{\sigma,\sigma^\prime} |_{\rm crossed} 
&=& 2 \left ( \frac{g_A}{f} \right )^2 {\bar u}_{\sigma^\prime}(Mv+
p^\prime)\,
\frac{S\cdot q \,\,S\cdot q^\prime}{v\cdot (p-q^\prime) + {\rm i} \epsilon}
u_\sigma(Mv+p) \nonumber\\
&&\nonumber\\
V_P^{I=\frac32}(q,q^\prime)_{\sigma,\sigma^\prime} |_{\rm contact}
&=&  \frac{1}{4f^2} \, {\bar u}_{\sigma^\prime}(Mv+
p^\prime)\, (v\cdot q + v\cdot q^\prime )
\,u_\sigma(Mv+p) \nonumber\\
&&\nonumber\\
D(k) &\approx & \frac{1}{k^2-m^2 + {\rm i}\epsilon} \qquad S(p) \approx
\frac{1}{v\cdot p + {\rm i} \epsilon} \label{eq:def_cross}
\end{eqnarray}
where the velocity $v$ is a time-like unit four-vector, in terms of
which the nucleon momenta can be written as $P-q\,(^\prime) = Mv +
p\,(^\prime)$ with $v\cdot p \,(^\prime) << M$. The EFT can be written
in terms of nucleon fields $N_v(x)$ with a definite velocity $v^\mu$
which are related to the original nucleon fields $\Psi (x) $ by
$N_v(x) = e^{{\rm i}\, M \not \, v \, v\cdot x }\Psi (x)$. On the
other hand, $S_\mu = \frac{{\rm i}}{2} \gamma_5 \sigma_{\mu\nu}v^\nu$
is defined in terms of the velocity and the Dirac
matrices\footnote{The following properties are satisfied~\protect\cite{BKM95}
\begin{equation}
S\cdot v = 0 \quad S^2 = -\frac34 \qquad \{ S_\mu, S_\nu\} = - \frac12
\left ( g_{\mu\nu} - v_\mu v_\nu \right ) \qquad [ S_\mu, S_\nu ] =
{\rm i}\, \epsilon_{\mu\nu\rho\alpha}v^\rho S^\alpha
\end{equation}
with $\epsilon_{0123} = +1$ .}. Finally, $f= 92.4$ MeV and $g_A=1.26$
are the pion weak decay constant and the vector axial coupling,
respectively. Note that because of isospin conservation, the direct Born
term does not contribute to the $I=\frac32$ {\it potential}. Likewise
due to angular momentum conservation the contact term does not
contribute either to the $P_{33}$ $ \pi N$ scattering and we will
ignore it from now on. Thus at lowest order, the BSE
(Eq.~(\ref{eq:bs})) for the $P_{33}$  channel reads:
\begin{eqnarray}
t_P^{I=\frac32}(q,q^\prime) &=& 2 \left ( \frac{g_A}{f} \right )^2
\,\frac{S\cdot q \,\,S\cdot q^\prime}{v\cdot (p-q^\prime)+{\rm i}\epsilon} 
+ 2 {\rm i} \left ( \frac{g_A}{f} \right )^2 \, 
\int \frac{d^4 k}{(2\pi)^4} \,t_P^{I=\frac32}(k,q^\prime) \times
\nonumber \\
&&\nonumber\\ 
&&   \frac{1}{k^2-m^2 +
{\rm i}\epsilon} \,\frac{1}{v\cdot P -M -v\cdot k + {\rm i} \epsilon} 
\,\frac{S\cdot q \,\,S\cdot k}
{v\cdot P - M -v\cdot(k+q)+{\rm i}\epsilon} \label{eq:bs_hb}
\end{eqnarray}
We look for solutions to the above equation, suggested by a one-loop
calculation, of the form
\begin{eqnarray}
t_P^{I=\frac32}(q,q^\prime) &=& A(v\cdot q, v\cdot q^\prime\,)\, S\cdot q
\,S\cdot q^\prime +  B (v\cdot q, v\cdot q^\prime\,) \,S\cdot q^\prime 
\,S\cdot q
\end{eqnarray}
where $A,B$ are functions of two variables to be determined. Note
that, as a simple one loop calculation shows, there appears a new
dependence ($S\cdot q^\prime \,S\cdot q$) not present in the
lowest order {\it potential}  $V^{I=\frac32}$. That is similar to what
happens in standard HBChPT~\cite{Mo98,FMS98}. The Eq.~(\ref{eq:bs_hb})
determines the functions $A,B$, which turn out to satisfy the
following integral equations
\begin{eqnarray}
 A(v\cdot q, v\cdot q^\prime\,) &=& 2 \left ( \frac{g_A}{f} \right )^2
\,\frac{1}{v\cdot P - M - v \cdot (q +q^\prime)+{\rm i}\epsilon } 
- \frac13 \left (
\frac{g_A}{f} \right )^2 {\rm i}\, \int \frac{d^4 k}{(2\pi)^4} 
   \frac{k^2-(v\cdot k)^2}{k^2-m^2 +
{\rm i}\epsilon}\nonumber\\ 
&&\nonumber\\
&\times&  \,\frac{1}{v\cdot P -M -v\cdot k + {\rm i} \epsilon} 
\,\frac{A(v\cdot k, v\cdot q^\prime \,)}
{v\cdot P - M -v\cdot(k+q)+{\rm i}\epsilon} \label{eq:a-off}\\
&&\nonumber\\
&&\nonumber\\
 B(v\cdot q, v\cdot q^\prime\,) &=&  - \frac16 \left (
\frac{g_A}{f} \right )^2 {\rm i}\, \int \frac{d^4 k}{(2\pi)^4} 
   \frac{k^2-(v\cdot k)^2}{k^2-m^2 +
{\rm i}\epsilon}\,\frac{1}{v\cdot P -M -v\cdot k + {\rm i} \epsilon}
\nonumber\\ 
&&\nonumber\\
&\times&   
\,\frac{A(v\cdot k, v\cdot q^\prime \,)-B(v\cdot k, v\cdot q^\prime \,)}
{v\cdot P - M -v\cdot(k+q)+{\rm i}\epsilon}
\label{eq:int}
\end{eqnarray}
The above set of linearly coupled  integral equations needs to be
regularized, solved and renormalized. This program is highly
non-trivial, and hence to draw solutions to these equations 
we will make use of the analytical structure of the involved
functions. 

\section{Elastic Unitarity and On-Shell Dispersion Relations.}

For on-shell scattering and at leading order in the $1/M$ expansion,
implicit in HBChPT formalism, LAB and CM systems coincide. Then, 
choosing $v^\mu = (1, \vec{0}) $ and taking into account that in the
infinitely heavy nucleon  limit $v\cdot q = v\cdot q^\prime =
v\cdot P -M = \sqrt{m^2 + \vec{q\,}^2_{CM}}  = (s-M^2 +
m^2)/2\sqrt s \equiv \omega $, the pion
energy, we find
\begin{equation}
f^{3/2}_{3/2\,,1} (\omega) = -\frac{1}{24 \pi}(\omega^2 - m^2) A(\omega) 
\end{equation}
where $A(\omega) \equiv A(\omega, \omega) $ for on-shell
scattering. The function $B$ enters in the $P_{3 1}$ pion-nucleon
scattering amplitude, and we will discuss it elsewhere~\cite{ej2001}.
Elastic unitarity, Eq.~(\ref{eq:uni_f}), applied to the function
$A(\omega)$ requires for $\omega > m$
\begin{equation}
{\rm Im} A^{-1}(\omega) = \frac{1}{24\pi}(\omega^2 -m^2)^\frac32
\label{eq:uni}
\end{equation}
From Eq.~(\ref{eq:a-off}), the on-shell function $A(\omega)$
satisfies, for $\omega > m$,
\begin{eqnarray}
 A(\omega) &=& - 2 \left ( \frac{g_A}{f} \right )^2
\,\frac{1}{\omega} + \frac13 \left (
\frac{g_A}{f} \right )^2 {\rm i}\, \int \frac{d^4 k}{(2\pi)^4} 
   \frac{k^2-(v\cdot k)^2}{k^2-m^2 +
{\rm i}\epsilon}\nonumber\\ 
&&\nonumber\\
&\times&  \,\frac{1}{\omega -v\cdot k + {\rm i} \epsilon} 
\,\frac{A(v\cdot k, \omega \,)}
{v\cdot k -{\rm i}\epsilon} \label{eq:a-on} 
\end{eqnarray}
\begin{figure}
\vspace{-8cm}
\hbox to\hsize{\hfill\epsfxsize=0.65\hsize
\epsffile[52 35 513 507]{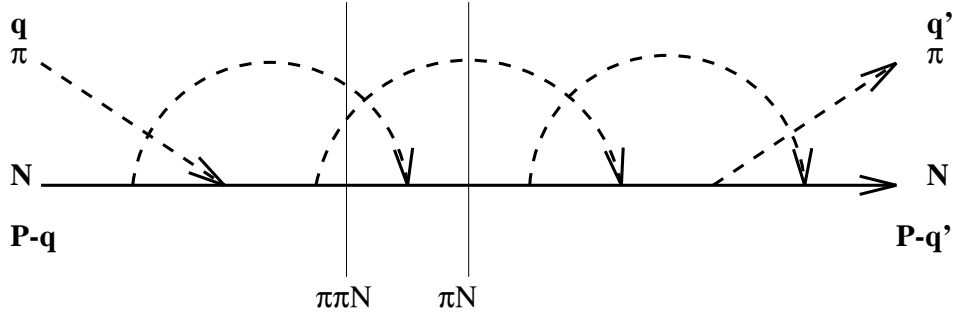}\hfill}
\vspace{1cm}
\caption{\footnotesize Diagrammatic representation of the BSE
equation obtained by iterating only the crossed term potential,
defined in Eq.~(\protect\ref{eq:def_cross}). Vertical lines represent
the $\pi N \to \pi N$ and $\pi N \to \pi \pi N$ cuts.}
\label{fig:cross-cuts}
\end{figure}
The function $A(\omega)$ above develops an imaginary part from two
difference sources: the elastic channel and the inelastic channel
corresponding to $\pi N \to \pi\pi N$ process, as can be seen in
Fig.~(\ref{fig:cross-cuts}). To obtain,
Eq.~(\ref{eq:uni}) we have completely neglected all possible
in-elasticities. We are going to proof that the solution of the
Eq.~(\ref{eq:a-on}) automatically satisfies the elastic unitarity
condition, given in Eq.~(\ref{eq:uni}), provided the pion energy,
$\omega$, is such that we are below the two pion threshold or,
equivalently, if the latter contribution to ${\rm Im} A(\omega)$ is
neglected. This can be easily
seen by standard operator methods by writing 
Eq.~(\ref{eq:a-off}) in an operator form 
\begin{eqnarray}
 A(\omega, \omega^\prime) &=& U(\omega, \omega^\prime) + 
\int d\nu \,U(\omega,\nu)\,G(\nu) \,A(\nu,\omega^\prime)  
\end{eqnarray}
with an obvious identification of the kernel $U$ and $G(\nu)$ given by
\begin{eqnarray}
G(\nu) &=& -\frac{i}{6} \int \frac{d^4 k}{(2\pi)^4} 
   \frac{k^2-(v\cdot k)^2}{k^2-m^2 +
{\rm i}\epsilon}\,\frac{1}{\alpha -v\cdot k + {\rm i} \epsilon} \delta
(\nu - v\cdot k) \label{eq:pro}
\end{eqnarray}
with $\alpha = v\cdot P -M$. If the inelastic $\pi N \to \pi\pi N$
channel is not considered, the potential $U(\omega, \omega^\prime)$ 
can be considered to be real, so that the corresponding 
i$\epsilon$ can be ignored.  Thus, one gets 
\begin{eqnarray}
A(\omega , \omega^\prime) -  A^*(\omega^\prime, \omega) &=& \int
d\nu\, A^*(\omega^\prime, \nu)\, 2{\rm i Im}G(\nu) A(\omega, \nu)
\label{eq:uni-a}
\end{eqnarray}
By direct application of Cutkosky's rules, we get from Eq.~(\ref{eq:pro})
\begin{eqnarray}
2{\rm i}{\rm Im}  G(\nu) &=& -\frac{i}{6} \int 
\frac{d^4 k}{(2\pi)^4} 
   \left [ k^2-(v\cdot k)^2\right] (-2{\rm i}\pi)^2\delta^+(k^2-m^2)\,
\delta(\alpha-v\cdot k)\, \delta
(\nu - v\cdot k) \nonumber \\
&=& \frac16 (m^2 - \alpha^2 ) \delta (\alpha -\nu) 2{\rm i} {\rm Im} 
J_0(\alpha)
\end{eqnarray}
where $\delta^+ (k^2 - m^2) = \theta (k^0) \delta (k^2-m^2) $ and
the one-loop function  
\begin{eqnarray}
J_0(\alpha) &=& -{\rm i} \int 
\frac{d^4 k}{(2\pi)^4} 
   \frac{1}{k^2-m^2+ {\rm i} \epsilon}\,
   \frac{1}{\alpha-v\cdot k + {\rm i} \epsilon}
\end{eqnarray}
has been defined. It is linearly divergent and hence two subtractions
are needed to make it convergent~\cite{BKM95}, 
\begin{eqnarray}
J_0(\alpha) &=& K_0 + K_1 \alpha  + \bar{J}_0 (\alpha)\\
&&\nonumber\\
\bar{J}_0 (\alpha) &=& -\frac{\sqrt{\alpha^2-m^2}}{4\pi^2} 
\{ {\rm arcosh} \frac{\alpha}{m} -{\rm i}
\pi\};  \qquad \alpha > m \label{eq:defjb}
\end{eqnarray}
where $K_0$ and $K_1$ are related to the divergent constants 
$J_0(0)$ and $J_0^\prime(0)$. Finally,
in the infinite nucleon mass limit, $\omega=\omega'=\alpha$, and
Eq.(\ref{eq:uni-a}) reduces to the unitarity condition of
Eq.(\ref{eq:uni}).

An exact solution of the BSE (\ref{eq:a-on}) contains both elastic and
two pion inelastic contributions, even though Eq.~(\ref{eq:a-on})
constitutes a lowest order approximation. This feature makes a
practical solution of that equation very difficult, mainly due to the
non-locality in the momentum variable of the exchanged nucleon. On the
other hand, we have seen that the solution of Eq.~(\ref{eq:a-on})
exactly satisfies the unitarity condition of Eq.(\ref{eq:uni}) when
in-elasticities are neglected. Besides, the inelastic contributions
start at ${\cal O}(1/f^4)$, but they are not complete because terms of
the same order have not even been included in the iterated {\it
potential} and {\it propagator} (vertex correction and self-energy
insertion). Thus, we propose to solve the BSE (\ref{eq:a-on}) in the
limit in which the inelastic $\pi N \to \pi\pi N$ channel does not
contribute to the imaginary part of $A(\omega)$, although this channel
may contribute to the real part of $A(\omega)$ as a subthreshold
effect. To do that, we make use of the analytical structure of
$A^{-1}(\omega)$ in the complex $\omega$ plane implied by the BSE to
the lowest order considered. Once that the two pion channel
contribution to ${\rm Im} A(\omega)$ is neglected, $A^{-1}(\omega)$
only has a right hand cut, with branch points in $\omega = m$ and
infinity , and a discontinuity through the cut determined by
Eq.~(\ref{eq:uni}),
\begin{equation}
A^{-1}(\omega + {\rm i } \epsilon) - A^{-1}(\omega - {\rm i } \epsilon) =  
2{\rm i Im } A(\omega + {\rm i } \epsilon)  = \frac{{\rm i }}{12\pi}
(\omega^2 - m^2)^\frac32 \label{eq:dis}
\end{equation}
On the other hand $A^{-1}(\omega)$ is not expected to have poles which
would correspond to zeros\footnote{Note that the kinematical zero at
$\omega = m$ has been factorized out} of $A(\omega)$ and thus of the
cross section. Under these circumstances and as demanded by the
asymtotic behaviour deduced from Eq.~(\ref{eq:dis}), , a four-times
subtracted dispersion relation for $A^{-1}(\omega)$ is needed, which
then reads\footnote{Note that, in this case, $A^{-1}(\omega)$ and $
(\omega^2-m^2)J_0(\omega)/6$ have the same analytical structure, and
therefore they can only differ by a polynomial in $\omega$.}  for
$\omega > m$:
\begin{eqnarray}
A^{-1}(\omega) &=& \frac{-f^2 \omega}{2 g_A^2} + P(\omega)
+ (\omega^2-m^2)\bar{J}_0(\omega)/6 \nonumber\\
&&\nonumber\\
P(\omega) &=& m^3\left ( c_0 + c_1
(\frac{\omega}{m}-1) + 
c_2 (\frac{\omega}{m}-1)^2 + c_3 (\frac{\omega}{m}-1)^3 \right
)\label{eq:ainv}
\end{eqnarray}
The coefficients $c_i$ might be expanded in powers of $1/f^2$. In the
former equation we have explicitly separated the lowest Born term
($-\omega f^2 /2 g_A^2$) which only affects to the linear coefficient
in $\omega$. Thus, we have $c_{0,1,2,3}= {\cal O}
(1)$ in the
infinite nucleon mass limit. We would like to make a few remarks:
\begin{itemize}
\item The structure of the solution, Eq.~(\ref{eq:ainv}), of
Eq.~(\ref{eq:a-on}), agrees with the findings of Ref.~\cite{ej2000},
where it was shown that in the case of ChPT in the meson-meson sector,
to get the on-shell amplitude, the off-shellness of the BSE could be
ignored, as long as a renormalized {\it potential} is iterated. With
this philosophy in mind, Eq.~(\ref{eq:a-on}) becomes
\begin{eqnarray}
A^{-1}(\omega) &=& H(\omega) - 
\frac{\omega f^2}{2g_A^2}  \nonumber \\
&&\nonumber \\
H(\omega) &=& \frac {\rm i}{6} \int 
\frac{d^4 k}{(2\pi)^4} 
   \frac{k^2- (v\cdot k)^2}{k^2-m^2+ {\rm i} \epsilon}\,
   \frac{1}{\omega-v\cdot k + {\rm i} \epsilon} \label{eq:ainv-tes}
\end{eqnarray}
The divergent Integral $H(\omega)$ is related to $J_0(\omega)$,
$H(\omega) = (\omega^2 - m^2)J_0(\omega)/6 + P_1(\omega)$, with
$P_1(\omega)$ a polynomial in the variable $\omega$ of degree one. The
above equation, is equivalent to Eq.~(\ref{eq:ainv}). Note, that in
this approximation, the nucleon propagator which enters in the
iterated {\it potential}, see Eq.~(\ref{eq:def_cross}), has been taken
out of the integral and hence it can never be put on the
mass-shell. This is the reason why the approximate solution given in
Eq.~(\ref{eq:ainv-tes}) does not contain the inelastic channel
$N\pi\pi$ contribution (see Fig.~\ref{fig:cross-cuts}) and thus it can
coincide with the solution of Eq.~(\ref{eq:ainv}), based only on the
elastic unitarity requirement.
 
\item If one is going to neglect explicitly the inelastic channel,
then only the nucleon propagator $1/(\omega -v\cdot k + {\rm i}
\epsilon)$ contributes to the elastic cut. This suggests to make a Taylor
expansion around $\omega$ in the integral of Eq.~(\ref{eq:a-on}), it
is to say
\begin{eqnarray}
\frac{k^2-(v\cdot k)^2}{k^2-m^2 +
{\rm i}\epsilon} \,\frac{A(v\cdot k, \omega \,)}
{v\cdot k -{\rm i}\epsilon} &=& \frac{k^2-\omega^2}{k^2-m^2 +
{\rm i}\epsilon} \,\frac{A(\omega)}
{\omega } + {\cal O}(v\cdot k -\omega)
\end{eqnarray} 
The leading order of this expansion yields again to Eq.~(\ref{eq:ainv}).

\end{itemize}

\section{Numerical Results}

To prove the usefulness of our main result, Eq.~(\ref{eq:ainv}), we
perform a direct $\chi^2 $ fit to the experimental phase shift in the
$P_{33}$ channel~\cite{AS95} from threshold up to $\sqrt{s}=$1500
MeV. We have assigned a 3\% uncertainty to the phase shift as
in~\cite{FMS98} plus a systematic error of one degree to the data
(similar treatments are followed in~\cite{MO2000}
and~\cite{NNPR2000}).

\begin{figure}

\begin{center}                                                                
\leavevmode
\epsfysize = 650pt
\makebox[0cm]{\epsfbox{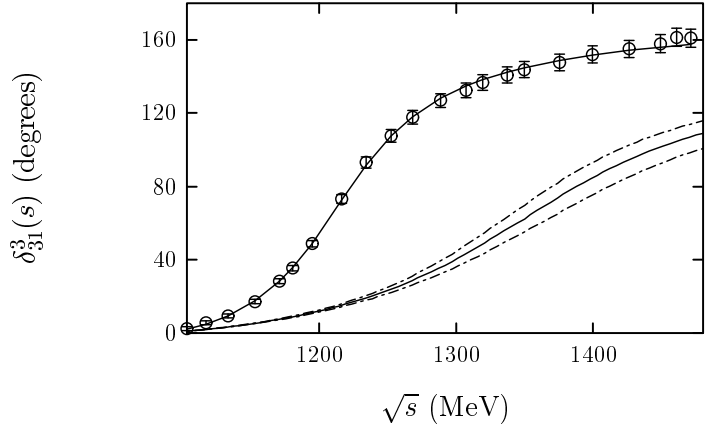}}
\end{center}
\vspace{-14.5cm}
\caption[pepe]{\footnotesize $P_{33}$ phase shifts as a function of
the total CM energy $\protect\sqrt s$. The upper solid line represents a
$\chi^2$-fit of the parameters of Eq.~(\protect\ref{eq:ainv}) to data
of Ref.~\protect\cite{AS95} (circles). Best fit parameters are given
in Eq.~(\protect\ref{eq:res}). The lower lines stand for the results
obtained with the parameters given in Eq.~(\protect\ref{eq:cesth}). 
Central values lead to the solid line, whereas the errors on
Eq.~(\protect\ref{eq:cesth}) lead to the dash-dotted lines.}
\label{fig:delta}
\end{figure}

The $\chi^2 $ fit yields the following numerical values for the parameters:
\begin{eqnarray}
c_0  = 0.045 \pm 0.021 &\quad&  c_1  = 0.29 \pm 0.08
\nonumber \\
&&\nonumber\\
c_2  = -0.17 \pm 0.09 &\quad & c_3 = 0.16 \pm 0.03 ,  
 \label{eq:res}
\end{eqnarray}
with $ \chi^2/{\rm d.o.f.} = 0.2 $. The description of the data from
threshold to the region well above the resonance is pretty good as can
be clearly seen in Fig.~\ref{fig:delta} (upper solid line).

On the other hand, the above coefficients can also be obtained from
the ${\cal O} (1/f^4)$ HBChPT pieces obtained in Refs.~\cite{Mo98}
and~\cite{FMS98}. Matching the BSE scattering amplitude to that
obtained in either of the above references, and in the infinite
nucleon mass limit, we find
\begin{eqnarray}
P(\omega) &=& \left. \frac{6\pi m^2 \omega^2}{(\omega^2-m^2) g_A^4}
\times  \left (m t^{(3,3)} - 4\pi \bar{J}_0(\omega) 
[t^{(1,1)}]^2  
 \right) \right|_{{\rm  Pol.~ of~ degree~ 3~ in~~} (\omega -m)}
\label{eq:polinoth} 
\end{eqnarray}
where the dimensionless amplitudes $t^{(3,3)}$ and $t^{(1,1)}$ are
defined in Ref.~\cite{NNPR2000}. They depend only on $\omega/m$, $g_A$
and the Low Energy Constants (LEC's):$\tilde{b}_1+\tilde{b}_2$,
$b_{16}-\tilde{b}_{15}$ and $b_{19}$ in the notation of
Ref.~\cite{Mo98}. The right hand side in Eq.~(\ref{eq:polinoth}) is
not a polynomial by itself, because it contains chiral logs stemming
from the left hand cut, and it has to be Taylor expanded around
$\omega=m$ to comply with the polynomial structure of
Eq.~(\ref{eq:ainv}). 

Using the values obtained in Ref.~\cite{FMS98} for the LEC's and
translated into the notation of ~\cite{Mo98} as shown 
in Table I of Ref.~\cite{NNPR2000}  (Set {\bf II} entry)
we find after Montecarlo propagation of errors, the lower solid and 
dash--dotted lines in 
Fig.~\ref{fig:delta} and estimates for the coefficients $c_{0,1,2,3}$
\begin{eqnarray}
c_0^{\rm th}   = 0.001 \pm 0.003 &\quad&  c_1^{\rm th}  = 0.038 \pm 0.006
\nonumber \\
&&\nonumber\\
c_2^{\rm th}  = 0.064 \pm 0.005 &\quad & c_3^{\rm th} = 0.036 \pm 0.002 
\label{eq:cesth}
\end{eqnarray}

Obviously, the structure of the Eq.~(\ref{eq:ainv}) which ensures
exact elastic unitarity, allows for a satisfactory fit to data, as
long as the contributions coming from 
two pion production are small and the left-hand-cut contribution can
be modeled, in the scattering region, by a polynomial of the third
degree in the variable $\omega-m$.

In the extreme static approximation, the HBChPT estimate of the
polynomial $P(\omega)$ provides the bulk of the resonant shape,
although it is far from describing accurately the data. Similar
conclusions, within the IAM scheme, were reached in
Ref.~\cite{NNPR2000}. There also, it was shown that the $1/M f^2$
corrections (NLO in the chiral expansion) are extremely important. 
This might explain the big discrepancies between the set of
parameters of Eqs.~(\ref{eq:res}) and~(\ref{eq:cesth}). Results in
Eq.~(\ref{eq:res}) are obtained from a best fit to data and
effectively incorporate, buried in the fitted parameters, all higher
order corrections, in particular the above mentioned $1/M f^2$ pieces.

\section{Conclusions}

In this paper we have dealt with the study of the $\Delta(1232)$
resonance as it arises in $\pi N$ scattering in the $P_{33}$
channel. We have used HBChPT as a perturbative guide to implement in a
sensible way the constraints imposed by chiral symmetry. The
Bethe-Salpeter equation provides, in addition, a workable scheme where
unitarity is exactly restored up to the level imposed by the
corresponding kernel and, moreover, allows for an identification of
the infinite set of diagrams being summed up. In the case of $\pi N$
scattering in the $P_{33}$ channel we have used HBChPT to lowest order
as input. The BSE is able, out of its divergence structure, to
generate some of the higher order counter-terms required in standard
HBChPT in the form of subtraction constants for the divergent
integrals. Actually, our BSE amplitude admits a very good fit to the
experimental phase shifts in terms of four subtraction constants.
Nevertheless, their fitted values differ, at least within estimated
errors, from those deduced from HBChPT. The discrepancy may be
understood because the NLO ($1/M f^2$) contributions
are comparable to the LO ($1/f^2$) ones, and hence the matching
procedure of the BSE solution to standard HBChPT amplitudes induces
large numerical corrections. This situation is not new and was already
encountered in previous studies based on the
IAM~\cite{NP2000,NNPR2000} which require a similar matching procedure
of unitarized amplitudes to the standard ones of HBChPT. As we see,
however, the amount of work required to describe the $\pi N $
scattering data within the BSE, and in particular, the $\Delta(1232)$
resonance is drastically smaller than that needed for the IAM. In
addition, as we pointed out in Ref.~\cite{ej2000} for the meson-meson
problem, one can search for systematic improvement and convergence in
a BSE framework. A full study of the $\pi N$ system along these lines
is left for future research~\cite{ej2001}.

\section*{Acknowledgments}
This research was supported by DGES PB98--1367 and by the Junta
de Andaluc\'{\i}a FQM0225.


\begin{thebibliography}{99}

\bibitem{gl84} J. Gasser and H. Leutwyler, Ann. of Phys., NY {\bf 158}
(1984) 142.


\bibitem{gss88} J. Gasser, M.E. Sainio and A. Svarc, Nucl. Phys. {\bf
B307} (1988) 779.

\bibitem{jm91} E. Jenkins and A. V. Manohar, Phys. Lett. {\bf B255}
(1991) 558. 

\bibitem{iw89} N. Isgur and M.B. Wise, Phys. Lett. {\bf B232} (1989)
113.

\bibitem{BL99} T. Becher and H. Leutwyler, Eur. Phys. Jour. {\bf C 9}
(1999) 643.

\bibitem{BK92} V. Bernard, N. Kaiser, J. Kambor and U. -G. Mei\ss ner, 
Nucl. Phys. {\bf B388} (1992) 315.  
\bibitem{BKM95} For a review see e.g. V. Bernard, N. Kaiser and 
U.-G. Mei\ss ner, Int. J. Mod. Phys. {\bf E4} (1995) 193 and 
references therein.
\bibitem{BKM97} V. Bernard, N. Kaiser and U.-G. Mei\ss ner,
Nucl. Phys. {\bf A615} (1997) 483.
\bibitem{Mo98} M. Mojzis, Eur. Phys. Jour.  {\bf C2} (1998) 181.
\bibitem{FMS98} N. Fettes, U.-G. Mei\ss ner and S. Steininger,
Nucl.Phys. {\bf A640} (1998) 199.
\bibitem{FM00} N. Fettes and U-G. Mei{\ss}ner, hep-ph/0002162.

\bibitem{ew88} T. Ericson and W. Weise, {\it Pions and Nuclei},
Clarendon Press, Oxford, 1988. 

\bibitem{MO2000} U.-G. Mei\ss ner and J.A. Oller, Nucl. Phys. {\bf
A673} (2000) 311.


\bibitem{NP2000} A. Gomez Nicola and J.R. Pelaez, Phys. Rev. {\bf
D 62} (2000) 017502.

\bibitem{NNPR2000} A. Gomez Nicola, J. Nieves, J.R. Pelaez and 
E. Ruiz Arriola, Phys. Lett. {\bf B486} (2000) 77.

\bibitem{BS51} H.A. Bethe and E.E. Salpeter, Phys. Rev. {\bf 82} (1951)
309; for a review see e.g. N. Nakanishi, Suppl. Prog. Theor. Phys.,
vol.{\bf 43} (1969) 1.

\bibitem{ej99} J. Nieves and E. Ruiz Arriola, Phys. Lett. {\bf B455}
(1999) 30. 

\bibitem{ej2000} J. Nieves and E. Ruiz Arriola, 
Nucl. Phys. {\bf A} in print, hep-ph/9907469.


\bibitem{ej2001} J. Nieves and E. Ruiz Arriola, work in progress. 



\bibitem{AS95} R.A. Arndt, I.I. Strakovsky, R.L. Workman, and M.M. Pavan,
Phys. Rev. {\bf C 52}, 2120 (1995). R.A. Arndt, {\it et. al.,}
nucl-th/9807087. SAID online-program (Virginia Tech Partial--Wave
Analysis Facility). Latest update, http://said.phys.vt.edu. 

%--------------







\end{thebibliography}
\end{document}